\documentclass[nohyperref]{article}

\usepackage{microtype}
\usepackage{graphicx}

\usepackage{subfigure}
\usepackage{booktabs} % for professional tables

\usepackage{hyperref}

\usepackage[accepted]{icml2022}

\usepackage{amsmath}
\usepackage{amssymb}
\usepackage{mathtools}
\usepackage{amsthm}
\usepackage[utf8]{inputenc}
\usepackage{graphicx}
\usepackage[capitalize,noabbrev]{cleveref}

\theoremstyle{plain}

\theoremstyle{definition}

\theoremstyle{remark}

\usepackage[textsize=tiny]{todonotes}

\DeclareRobustCommand{\rchi}{{\mathpalette\irchi\relax}}
\newcommand{\irchi}[2]{\raisebox{\depth}{$#1\chi$}}

\icmltitlerunning{TNT: Vision Transformer for Turbulence Simulations}

\begin{document}

\twocolumn[
\icmltitle{TNT: Vision Transformer for Turbulence Simulations}
%Turbulence Neural Transformer
% IDEA: Vision transformer for (turbulence) simulations? 

% It is OKAY to include author information, even for blind
% submissions: the style file will automatically remove it for you
% unless you've provided the [accepted] option to the icml2022
% package.

% List of affiliations: The first argument should be a (short)
% identifier you will use later to specify author affiliations
% Academic affiliations should list Department, University, City, Region, Country
% Industry affiliations should list Company, City, Region, Country

% You can specify symbols, otherwise they are numbered in order.
% Ideally, you should not use this facility. Affiliations will be numbered
% in order of appearance and this is the preferred way.
\icmlsetsymbol{equal}{*}

\begin{icmlauthorlist}
\icmlauthor{Yuchen Dang}{NYUCDS}
\icmlauthor{Zheyuan Hu}{NYUCDS}
\icmlauthor{Miles Cranmer}{Princeton}
\icmlauthor{Michael Eickenberg}{CCM}
\icmlauthor{Shirley Ho}{CCA,DASPrinceton,CMU}
\end{icmlauthorlist}

\icmlaffiliation{NYUCDS}{Center for Data Science, New York University, New York, NY, USA}
\icmlaffiliation{Princeton}{Department of Astrophysical Sciences, Princeton University, Princeton, NJ, USA}
\icmlaffiliation{CCM}{Center for Computational Mathematics, Flatiron Institute, 162 5th Ave, New York, NY, 10010, USA}
\icmlaffiliation{CCA}{Center for Computational Astrophysics, Flatiron Institute, 162 5th Ave, New York, NY, 10010, USA}
\icmlaffiliation{DASPrinceton}{Department of Astrophysical Sciences, Princeton University, Peyton Hall, Princeton, NJ, 08544, USA}
\icmlaffiliation{CMU}{Department of Physics, Carnegie Mellon University, Pittsburgh, PA 15213, USA}
\icmlcorrespondingauthor{Yuchen Dang}{yd1008@nyu.edu}
\icmlcorrespondingauthor{Zheyuan Hu}{zh2095@nyu.edu}

% You may provide any keywords that you
% find helpful for describing your paper; these are used to populate
% the "keywords" metadata in the PDF but will not be shown in the document
\icmlkeywords{Machine Learning, ICML}

\vskip 0.3in
]

% this must go after the closing bracket ] following \twocolumn[ ...

% This command actually creates the footnote in the first column
% listing the affiliations and the copyright notice.
% The command takes one argument, which is text to display at the start of the footnote.
% The \icmlEqualContribution command is standard text for equal contribution.
% Remove it (just {}) if you do not need this facility.

%\printAffiliationsAndNotice{}  % leave blank if no need to mention equal contribution
% \printAffiliationsAndNotice{\icmlEqualContribution} % otherwise use the standard text.
\printAffiliationsAndNotice{}

\begin{abstract}
Turbulence is notoriously difficult to model due to its multi-scale
%d complexity 
nature
and 
%vulnerability
sensitivity
to small perturbations. Classical solvers of turbulence simulation generally operate on finer grids and are computationally inefficient. In this paper, we propose the \textit{Turbulence Neural Transformer} (TNT), which is a learned simulator based on the transformer architecture, to predict turbulent dynamics on coarsened grids. TNT extends the positional embeddings of vanilla transformers to a spatiotemporal setting to learn the representation in the 3D time-series domain, and applies Temporal Mutual Self-Attention (TMSA), 
%that
which
captures adjacent dependencies, to extract deep and dynamic features. TNT is capable of generating comparatively long-range predictions stably and accurately, and we show that TNT outperforms the state-of-the-art U-net simulator on several metrics. We also test the model 
%on its 
performance with 
%multiple
different
components removed and
evaluate
robustness 
%against 
to
different initial conditions. Although more 
%future 
experiments are needed, we conclude that TNT has %the 
great potential to outperform existing solvers and generalize 
%on
to
additional simulation datasets.
\end{abstract}

\section{Introduction}
Many areas of the 
%world 
science and engineering
rely heavily on turbulent dynamics. 
%Turbulence
Turbulent
flow simulations are widely employed in a variety of fields such as
aerospace \cite{mishra2019uncertainty}, weather \cite{mirocha2014resolved}, combustion \cite{pitsch2006large}, and medicine \cite{ha2017estimating}.
%Need to pick older papers to cite here?
Due to the complexity of temporal and spatial scales and the chaotic nature of turbulence \cite{berera2018chaotic,duraisamy2019turbulence}, in which tiny perturbations in 
%beginning 
initial
conditions can result in huge changes in outcomes, turbulence prediction is a difficult task despite its extensive uses.

In recent decades, numerical solvers for simulating large and complicated turbulent dynamics have been developed \cite{kim1999unsteady,kang2011high}.
However, these solvers rely heavily on massive computational resources and high spatial and temporal resolutions. 
More recently, solvers based on learned Machine Learning (ML) models have emerged, including 
%learned PDE solvers \cite{long2019pde,li2020fourier}, 
Graph Neural Networks (GNN) \cite{sanchez2020learning,li2020neural,pfaff2020learning} and Convolutional Neural Networks (CNN) \cite{pathak2020using,stachenfeld2021learned,li2020fourier,wang2020towards}.
%Fourier neural operater cNN and pde
ML methods are feasible and promising ways to build more connections between existing simulation results with real-world data and to automatically identify dependencies which are not explained by current physics and mathematical equations \cite{beck2021perspective}.

Our contribution is to introduce 
%a
the
\textit{
Turbulence Neural Transformer 
}
(TNT)
to learn turbulent dynamics on coarse grids based on the attention mechanism as introduced in the Transformer model \citep{vaswani2017attention}. TNT uses the Temporal Mutual Self Attention (TMSA) block \cite{liang2022vrt} to extract dependencies from an input window and generate rolling predictions with an encoder-decoder Transformer. It does not require any domain-specific expertise and is capable of producing more consistent and accurate predictions than the compared benchmark across %a 
relatively long and unevenly 
%spaced 
sampled
temporal ranges. 
\section{Related work}
%Transformer

\begin{figure*}[h]
\vskip 0.1in
\begin{center}
\centerline{\includegraphics[width=1.65\columnwidth]{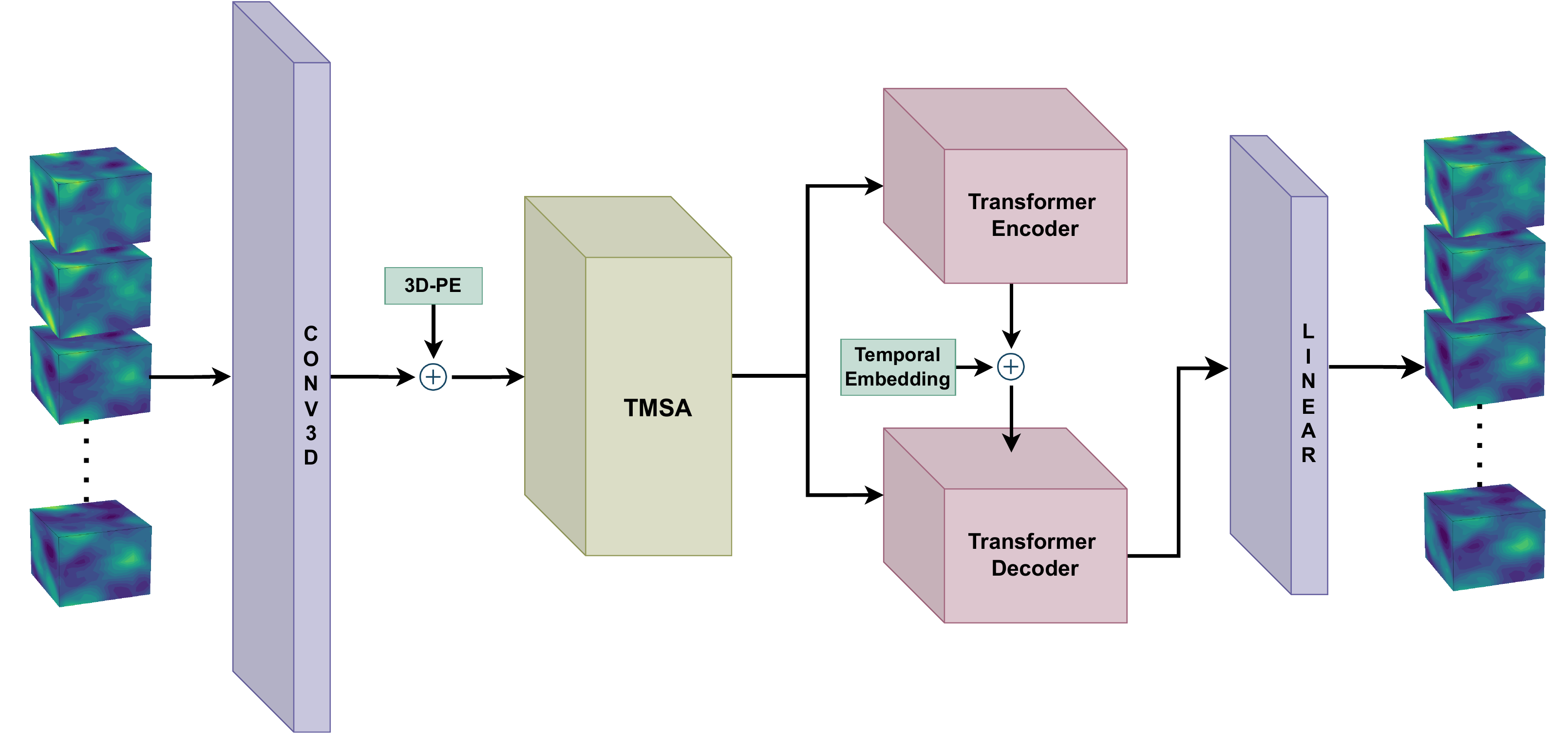}}
\caption{Framework of proposed Turbulence Neural Transformer (TNT) model. It takes a window of $16^3$ CompDecay-3D volumes as inputs and outputs a same-sized shifted window as predictions. TMSA and 3D-PE incorporate spatial information and temporal information is infused with temporal embedding.}
\label{fig:model-architecture}
\end{center}
\vskip -0.1in
\end{figure*}

Transformer networks \cite{vaswani2017attention}
%has
have
been the state-of-the-art model for the majority NLP tasks since 
%its 
their
introduction, along with other subsequent models based on the attention mechanism, such as BERT \cite{devlin2018bert} and  RoBERTa \cite{liu2019roberta}. Recent studies have demonstrated the immense potential of using attention for multiple tasks in Computer Vision (CV) including 
%image 
object 
recognition in images \cite{dosovitskiy2020image,liu2021swin} and video restoration \cite{liang2022vrt,Kim_2018_ECCV}.  
%Furthermore, transformer has gained increasing popularity in time series tasks such as in univariate influenza-like illness forecasting \cite{wu2020deep} and multivariate long-sequence forecasting \cite{zhou2021informer}.

There are several previous studies on learned ML simulators for turbulence dynamics. \citet{li2020fourier} built an efficient Fourier Neural Operator to solve PDEs including Navier-Stokes. \citet{wang2020towards} employed learnt spectral filters and U-net to construct a Turbulence-Flow Net for turbulent predictions. \citet{stachenfeld2021learned} implemented a general-purposed Dilated ResNet architecture to learn turbulent dynamics in a supervised setting without integrating specific domain knowledge. 
Our work is greatly inspired by studies of \citet{dosovitskiy2020image}, \citet{liang2022vrt}, and \citet{stachenfeld2021learned}. We take advantage of 
the recent advancements using
Transformers
%'s advancements 
in CV and apply those 
%on
to
modeling turbulence.
%Prev Simulations
%Video transformers?

\section{Data}
Given a
%n 
temporally
unevenly 
%spaced 
sampled
dataset $X \in \mathbb{R}^{N\times H\times W\times D\times C}$, where $N$ is the total number of 3-dimensional 
%blocks
volumes
ordered as a temporal sequence%
, $H,W,D$ are the grid size on the three spatial axes, and $C$ is the number of features. Denote $X_{i}$ as the $i$th 
%block
volume%
, $\tau_i$ as the timestamp of $i$th 
volume%
%block
, $\Delta \tau_i$ as the interval between the $\tau_i$ and $\tau_{i+1}$, and $x,y,z$ as the spatial coordinates each ranging from $0$ to the size of the corresponding spatial axes.

Data in this study are obtained from Athena++ \cite{Stone2020}, a state-of-the-art tool for astrophysical simulations. Specifically, we examine the 3D Compressible Decaying Turbulence (CompDecay-3D) with unevenly spaced temporal axis where $\tau$ ranges from $0$ to $1$. There are a total of 160 simulated 
%blocks 
volumes
with grid 
%resolutions of 
size
$16^3$. At each grid point, there are five features, namely gas density $\rho$, velocity vectors on the three spatial axes $v_x,v_y,v_z$, and gas pressure $P$. 
The goal is to predict the changes in densities $\Delta\rho_i$ between the current and the next
%block
volume
given a current 
volume
%block
$X_{i}$ and a sequence of future timestamps $[\tau_{i+1},\tau_{i+2},\dots]$, with $\upsilon_x,\upsilon_y,\upsilon_z,P$ as auxiliary features.
There are two main challenges associated with CompDecay-3D. 
First, turbulence is a multi-scale physics phenomenon that necessitates extensive multi-scale computations. Given the coarse setting, it is particularly difficult because there is less available information from which the model may learn. 
%Moreover, the compressibility complicates forecasts by allowing densities to change during flows.
Second, the time increment $\Delta t$
%s vary 
varies
during the simulation, ranging from $4.3$ to $9.3$ milliseconds at roughly increasing temporal intervals throughout the simulation. Because of this inconsistency, the correlation between timestamps and turbulence dynamics needs to be adequately characterized.

\section{Model}
TNT is an attention-based model that learns the mapping $\boldsymbol{\rchi} \xrightarrow{} \boldsymbol{\rchi}$ where $\boldsymbol{\rchi} \subseteq \mathbb{R}^5$ and $X \in \boldsymbol{\rchi}$. Specifically, it takes a temporal window of 
%blocks 
volumes
$[X_i,X_{i+1},\dots,X_{i+w}]$ as inputs and output a shifted window of 
volumes
%blocks
$[\Tilde{X}_{i+1},\Tilde{X}_{i+2},\dots,\Tilde{X}_{i+w+1}]$ as predictions, where $w$ is the window size. Figure \ref{fig:model-architecture} illustrates the architecture of TNT.

\subsection{3D Positional Encoding}
With a simple modification to the position encoding technique proposed in the transformer paper \cite{vaswani2017attention}, we develop a 3D positional embedding (3D-PE). Three sets of 1D positional encodings are made separately on three spatial axes, each having one-third of the latent dimension size $d$. The final 3D-PE is constructed by concatenating three 1D encodings. Each grid point within the same volume receiving a unique encoding, whereas grid points with identical spatial coordinates at different timestamps receive identical encodings.

\subsection{Temporal Embedding}
Due to the unevenness on the temporal axis, instead of using fixed representations such as from a Fourier transformation of time values, we adopt a temporal embedding method named Time2Vec \cite{kazemi2019time2vec} that uses learned representations as the embedding values. 
%Temporal embedding is added to the input in later layers in the Transformer's decoder. %WHY? Need justifications
For a given scalar of time $\tau$ , Time2Vec of $\tau$, denoted as $\vec{t2v}(\tau)$, is a vector of size k + 1 defined as follows:
\begin{equation}
  \vec{t2v}(\tau)[i] =
    \begin{cases}
      \omega_i \tau + \varphi_i & \textbf{if $i=0$}\\
      \digamma(\omega_i \tau + \varphi_i) & \textbf{if $1 \le i \le k$} \label{eq:te}
    \end{cases}       
\end{equation}
where $\vec{t2v}(\tau)[i]$is the ith element of $\vec{t2v}(\tau)$, $\digamma$ is a sine activation function, and $\omega$ is and $\varphi$ is are learnable parameters indicating correspondingly frequency and phase-shift.

% \subsection{Feature extraction}
% The first layer of the TNT is a simple 3D convolutional layer (CONV3D). It retrieves shallow features from each volume in the input window, and all volumes share the same parameters in CONV3D. 
% %The 3D Temporal Mutual Self Attention (TMSA) block is then utilized to extract deeper and more dynamic features.

\begin{figure}[t]
\vskip 0.1in
\begin{center}
\centerline{\includegraphics[width=\columnwidth]{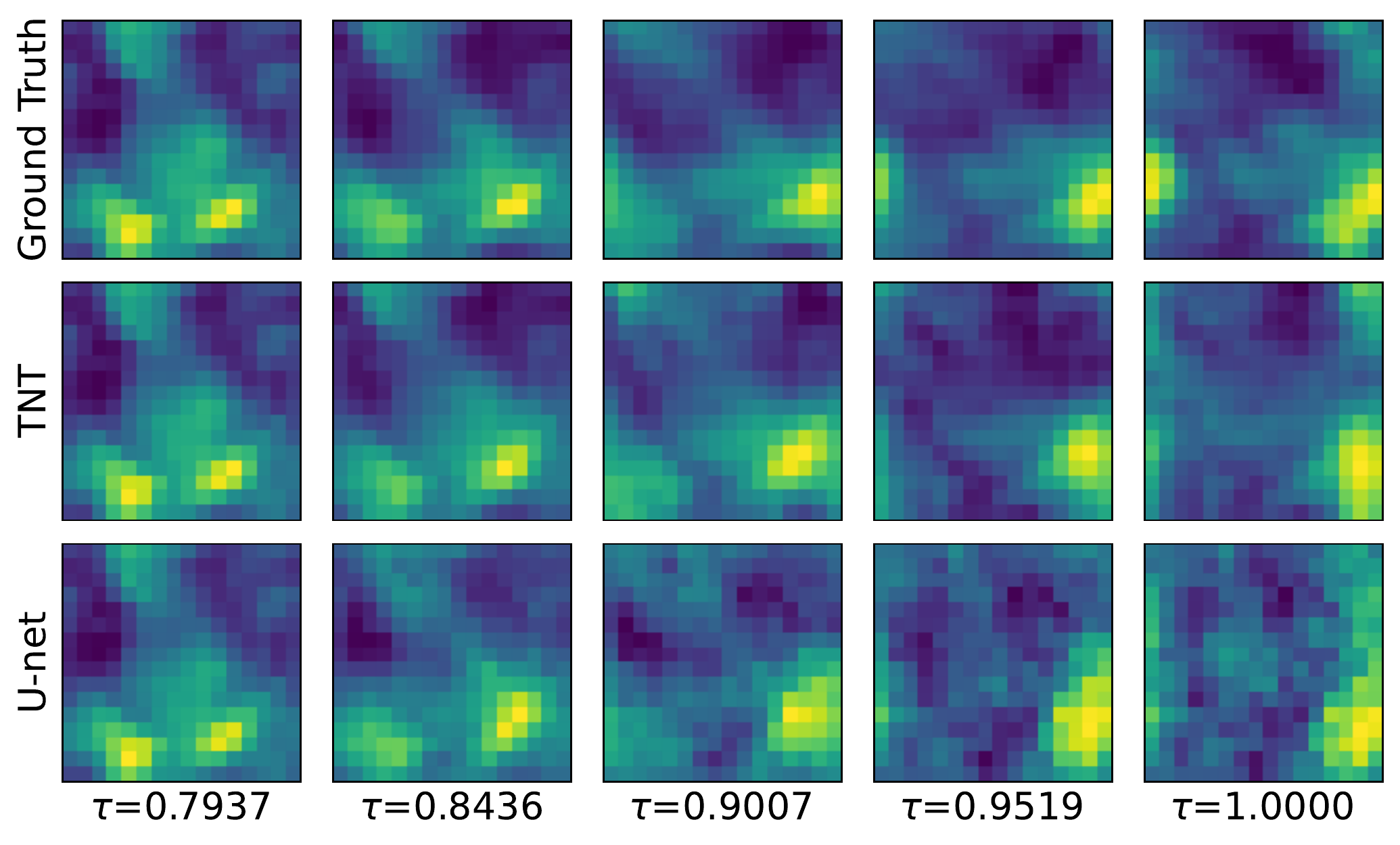}}
\caption{Visualization on predictions of U-net and TNT vs ground truth at five different timestamps from the beginning, $\tau=0.7937$, to the end, $\tau=1.0$, of the testing set. Each slice is taken at the middle of the volume across the third spatial axis $z$.}
\label{fig:slices_across_time}
\end{center}
\vskip -0.1in
\end{figure}

\subsection{TMSA}
Temporal Mutual Self-Attention is a feature alignment technique originally proposed, in Video Restoration Transformer (VRT)\cite{liang2022vrt} paper, to extract interactions between adjacent frames in a video.
% In one TMSA layer, a sequence of frames are partitioned into non-overlapping windows each containing two frames. Each set of two-frame window is then partitioned again into pair of non-overlapping patches, and mutual attention is used to warp each pair of patches towards each other and self attention is utilized on the concatenation of two patches to preserve information. The authors further shift the sequence by 1 frame in every other layer in a TMSA block with $n$ layers such that one frame in the final layer can comprise information up to $2(n-1)$ frames.
%add equation to explain？

We modified the TMSA to make it compatible with 3D input and the periodic boundary condition when input windows are shifted. Each layer in the 3D-TMSA block takes a window of volumes and partitions it into non-overlapping two-volumes windows which are further partitioned into pairs of smaller patches. Following the same idea as in VRT, we apply 3D-TMSA on each pair of patches and shift the entire window of volumes on the three spatial axes and the temporal axis.
%Did x1 x2 x3 already defined?
As a result, the output of the 3D-TMSA block provides information on both temporal and spatial interaction within an input window of volumes.
%CNNs could not capture dependencies along the temporal axis nor are they inherently informed of the positional information.

\subsection{Transformer}
We follow the same architecture of the vanilla Transformer \cite{vaswani2017attention} with minor modifications. Transformer's input volumes are partitioned into patches in the same manner as TMSA. Input to the decoder is a right-shifted window of the encoder's input along the temporal axis with the last volume filled with zeros. The Transformer's output is then fed into a simple linear layer that generates predictions.

\subsection{Boundary condition}
Due to the periodic boundary condition in the dataset, all the methods used in this study have been specifically configured or modified to capture the correlations across the boundaries.

\section{Experiments}

\subsection{Benchmark}
We compare TNT's performance with a 3D version of U-net \cite{ronneberger2015u} where code is modified from ELEKTRONN team's GitHub \cite{3d-unet-github}. U-net has been a popular benchmark and shown competitive performance in numerous previous works \cite{stachenfeld2021learned, li2020fourier, pathak2020using}. Our implementation of U-net closely follows that in \citet{stachenfeld2021learned} on choices of parameters and addition of training noise.

\subsection{Evaluation Metrics}
In this specific task, the criterion of the model performance is its ability to stably predict the turbulence density over time. Therefore, we selected Root Mean Squared Error (RMSE), Mean Absolute Error (MAE), R-Squared ($\text{R}^2$), and Explained Variance (EV) as our evaluation metrics. Detailed description and mathematical defination are in Appendix \ref{sec:B}. A better performance is indicated by smaller RMSE \& MAE and larger $\text{R}^2$ \& EV. 

\subsection{Training and fine-tuning} \label{sec:fine-tuning}
We split the CompDecay-3D dataset into training (60\%), validation (20\%), and testing (20\%) sets chronologically on the temporal axis. Predictions are made in a rolling manner where previous predictions are used in generating subsequent ones and each rollout step consist of prediction on the entire $16^3$ volume.
For parameter optimization, we employ a mean average error loss and an early stopping to prevent overfitting.
We use 4 RTX8000 GPUs for all fine-tuning. The best set of hyperparameters is selected based on performance on the validation split.
During inference, training is conducted on the combined training and validation sets using the chosen hyperparameters to generate predictions on the testing set.
%Need to describe tuning ranges of all hyperparameters?

\section{Results}
\begin{figure}[h]
\vskip 0.1in
\begin{center}
\centerline{\includegraphics[width=\columnwidth]{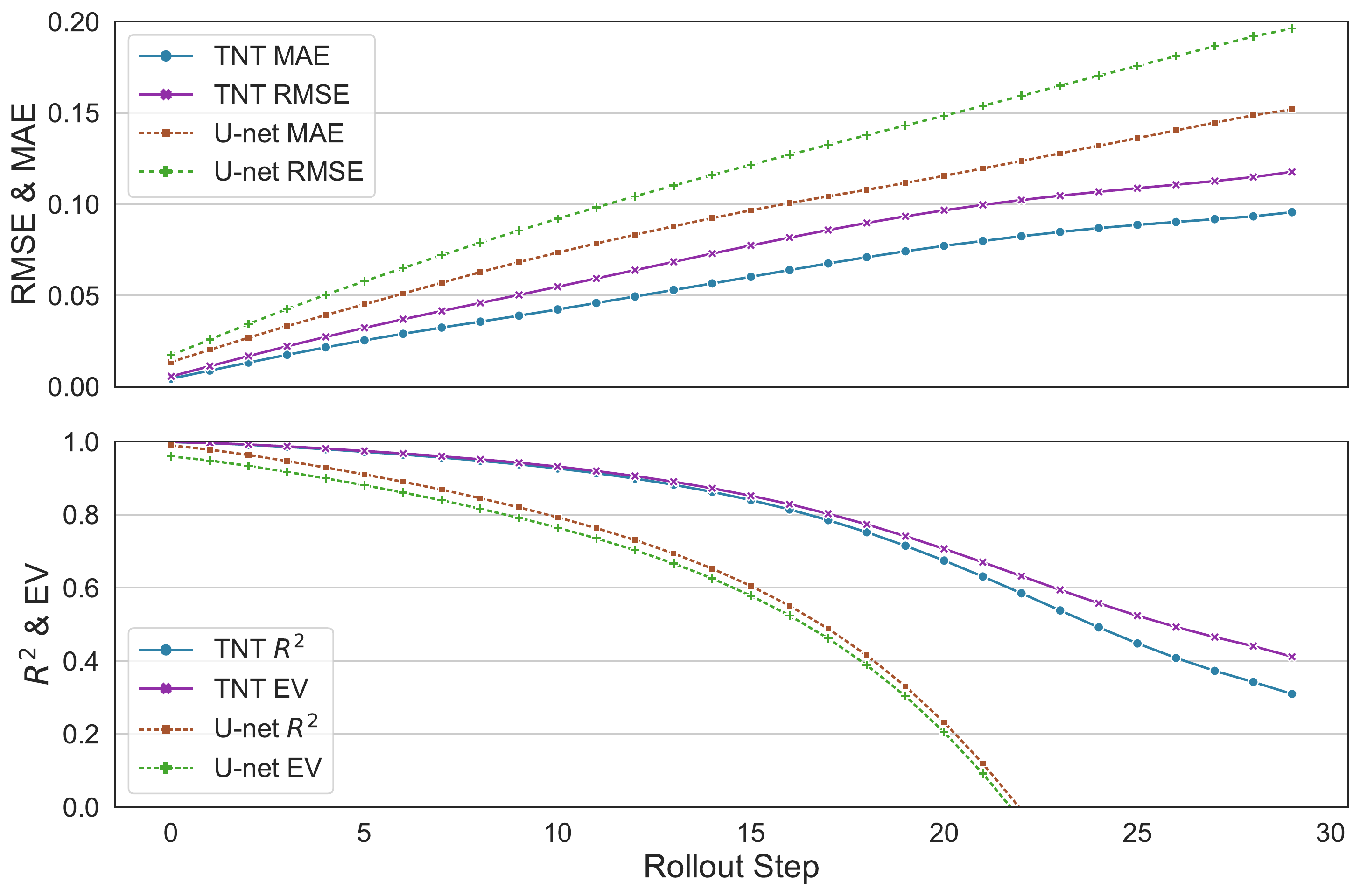}}
\caption{Comparison of performance between our proposed model TNT and U-net vs. number of rollout steps. TNT shows higher performance in all four evaluated metrics.}%The four metrics evaluated are Mean Averaged Error (MAE), Root Mean Squared Error (RMSE), R-squared ($\text{R}^2$), and explained variance (EV).
\label{fig:transformer-unet}
\end{center}
\vskip -0.1in
\end{figure}

We plot the progression of different metrics vs. number of rollout steps for both U-net and our proposed model TNT in Figure \ref{fig:transformer-unet}. As a result of rolling prediction, errors from previous predictions are carried over to subsequent ones. As observed in the plots, both models show the accumulation of errors with longer rollout steps. Throughout the inference phases, TNT outperforms U-net with smaller initial errors and slower degradation in performance. As illustrated in Figure \ref{fig:slices_across_time}, TNT generates stable and accurate predictions after 30 rollout steps, whereas U-net's predictions are noisy and less accurate. It is also observed in Table \ref{table:testset-result} that TNT performs better in all metrics with improvements of 35\% in RMSE, 35\% in MSE, 48\% in $\text{R}^2$, and 52\% in EV. TNT can effectively capture the multi-scaled dependencies in CompDecay-3D whereas U-net fails to generate predictions with longer rollout steps. Ablation studies and results on all features can be found in the Appendix \ref{sec:result-ablation} and \ref{sec:result-all-feature}.

\begin{table}[h]
\caption{Performance of TNT and U-net on the testing data}
\label{table:testset-result}
\vskip 0.1in
\begin{center}
\begin{small}
\begin{sc}
\begin{tabular}{lcccr}
\toprule
Model & RMSE & MAE & $\text{R}^2$ & EV \\
\midrule
\textbf{TNT} & 0.0782  & 0.0561 & 0.8080 & 0.8239 \\
U-net & 0.1208 & 0.0864 & 0.5401 & 0.5410 \\
\bottomrule
\end{tabular}
\end{sc}
\end{small}
\end{center}
\vskip -0.1in
\end{table}

\begin{figure}[h]
\vskip 0.1in
\begin{center}
\centerline{\includegraphics[width=\columnwidth]{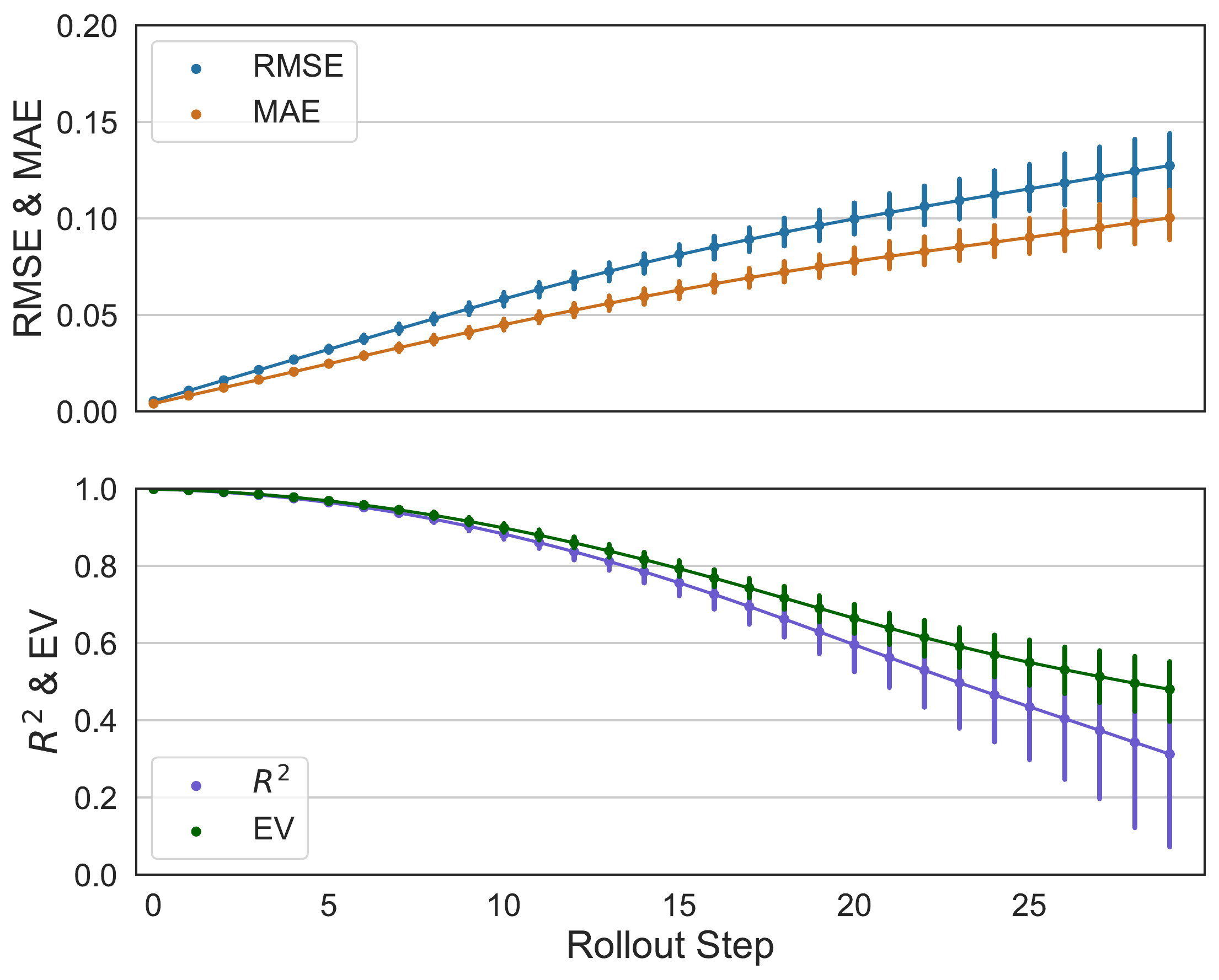}}
\caption{Generalization on 10 additional initial conditions. TNT is simply re-trained without further fine-tuning. No significant difference in performance from the primary result is observed acorss different initial conditions. }
\label{fig:seed_result}
\end{center}
\vskip -0.1in
\end{figure}

\subsection{Generalization on different initial conditions}

We also conduct experiments with ten distinct initial conditions, as slight variations in initial conditions can result in drastically different outcomes. Without further fine-tuning, we simply re-train the model with the same set of hyperparameters using the combined training and validation sets described in Section \ref{sec:fine-tuning}. Figure \ref{fig:seed_result} shows the line plot with error bars for all ten trails. There is no large deviance from the performance observed in Figure \ref{fig:transformer-unet}. Since the exact same tuned parameters are used throughout all experiments, there is sufficient evidence to support our claim that TNT, with simple re-training, can generalize to all potential simulations in CompDecay-3D.

\section{Conclusions}
Turbulence prediction is a difficult task due to the complexities of temporal and spatial scales. 
We conduct experiments on predicting features in CompDecay-3D with 11 different initial conditions on coarse grids of $16^3$. 
Our proposed model, TNT, shows significantly lower RMSE and MAE and higher $\text{R}^2$ and EV than a U-net. Although TNT is robust to initial conditions in CompDecay-3D, additional experiments are needed to demonstrate its ability on generalizing to larger grid resolutions and longer rollout steps.
We further show the importance of major components in our architecture in ablation studies in Appendix \ref{sec:result-ablation}, where significant drops in performance are observed with each component removed. 
The experimental results suggest that using attention to describe chaotic and complex simulations is feasible and effective.  The two attention-based layers, TMSA and Transformer, can aid in the capture of spatial dependencies, while temporal embedding aids in the capture of temporal dependencies in rollout. 
Our proposed model shows great potential in generalizing well to a larger variety of simulations for which additional experiments are required.

\clearpage
\newpage
\bibliography{reference}

\begin{thebibliography}{27}
\providecommand{\natexlab}[1]{#1}
\providecommand{\url}[1]{\texttt{#1}}
\expandafter\ifx\csname urlstyle\endcsname\relax
  \providecommand{\doi}[1]{doi: #1}\else
  \providecommand{\doi}{doi: \begingroup \urlstyle{rm}\Url}\fi

\bibitem[Beck \& Kurz(2021)Beck and Kurz]{beck2021perspective}
Beck, A. and Kurz, M.
\newblock A perspective on machine learning methods in turbulence modeling.
\newblock \emph{GAMM-Mitteilungen}, 44\penalty0 (1):\penalty0 e202100002, 2021.

\bibitem[Berera \& Ho(2018)Berera and Ho]{berera2018chaotic}
Berera, A. and Ho, R.~D.
\newblock Chaotic properties of a turbulent isotropic fluid.
\newblock \emph{Physical review letters}, 120\penalty0 (2):\penalty0 024101,
  2018.

\bibitem[Devlin et~al.(2018)Devlin, Chang, Lee, and Toutanova]{devlin2018bert}
Devlin, J., Chang, M.-W., Lee, K., and Toutanova, K.
\newblock Bert: Pre-training of deep bidirectional transformers for language
  understanding.
\newblock \emph{arXiv preprint arXiv:1810.04805}, 2018.

\bibitem[Dosovitskiy et~al.(2020)Dosovitskiy, Beyer, Kolesnikov, Weissenborn,
  Zhai, Unterthiner, Dehghani, Minderer, Heigold, Gelly,
  et~al.]{dosovitskiy2020image}
Dosovitskiy, A., Beyer, L., Kolesnikov, A., Weissenborn, D., Zhai, X.,
  Unterthiner, T., Dehghani, M., Minderer, M., Heigold, G., Gelly, S., et~al.
\newblock An image is worth 16x16 words: Transformers for image recognition at
  scale.
\newblock \emph{arXiv preprint arXiv:2010.11929}, 2020.

\bibitem[Duraisamy et~al.(2019)Duraisamy, Iaccarino, and
  Xiao]{duraisamy2019turbulence}
Duraisamy, K., Iaccarino, G., and Xiao, H.
\newblock Turbulence modeling in the age of data.
\newblock \emph{Annual Review of Fluid Mechanics}, 51:\penalty0 357--377, 2019.

\bibitem[Ha et~al.(2017)Ha, Lantz, Ziegler, Casas, Karlsson, Dyverfeldt, and
  Ebbers]{ha2017estimating}
Ha, H., Lantz, J., Ziegler, M., Casas, B., Karlsson, M., Dyverfeldt, P., and
  Ebbers, T.
\newblock Estimating the irreversible pressure drop across a stenosis by
  quantifying turbulence production using 4d flow mri.
\newblock \emph{Scientific reports}, 7\penalty0 (1):\penalty0 1--14, 2017.

\bibitem[Kang et~al.(2011)Kang, Lightbody, Hill, and
  Sotiropoulos]{kang2011high}
Kang, S., Lightbody, A., Hill, C., and Sotiropoulos, F.
\newblock High-resolution numerical simulation of turbulence in natural
  waterways.
\newblock \emph{Advances in Water Resources}, 34\penalty0 (1):\penalty0
  98--113, 2011.

\bibitem[Kazemi et~al.(2019)Kazemi, Goel, Eghbali, Ramanan, Sahota, Thakur, Wu,
  Smyth, Poupart, and Brubaker]{kazemi2019time2vec}
Kazemi, S.~M., Goel, R., Eghbali, S., Ramanan, J., Sahota, J., Thakur, S., Wu,
  S., Smyth, C., Poupart, P., and Brubaker, M.
\newblock Time2vec: Learning a vector representation of time.
\newblock \emph{arXiv preprint arXiv:1907.05321}, 2019.

\bibitem[Kim et~al.(2018)Kim, Sajjadi, Hirsch, and Scholkopf]{Kim_2018_ECCV}
Kim, T.~H., Sajjadi, M. S.~M., Hirsch, M., and Scholkopf, B.
\newblock Spatio-temporal transformer network for video restoration.
\newblock In \emph{Proceedings of the European Conference on Computer Vision
  (ECCV)}, September 2018.

\bibitem[Kim \& Menon(1999)Kim and Menon]{kim1999unsteady}
Kim, W.-W. and Menon, S.
\newblock An unsteady incompressible navier--stokes solver for large eddy
  simulation of turbulent flows.
\newblock \emph{International Journal for Numerical Methods in Fluids},
  31\penalty0 (6):\penalty0 983--1017, 1999.

\bibitem[Kornfeld et~al.(2018)Kornfeld, Drawitsch, Nguyen, Pfeiler, Schubert,
  Dorkenwald, and Killinger]{3d-unet-github}
Kornfeld, J., Drawitsch, M., Nguyen, M.-T., Pfeiler, N., Schubert, P.,
  Dorkenwald, S., and Killinger, M.
\newblock elektronn3.
\newblock \url{https://github.com/ELEKTRONN/elektronn3}, 2018.

\bibitem[Li et~al.(2020{\natexlab{a}})Li, Kovachki, Azizzadenesheli, Liu,
  Bhattacharya, Stuart, and Anandkumar]{li2020fourier}
Li, Z., Kovachki, N., Azizzadenesheli, K., Liu, B., Bhattacharya, K., Stuart,
  A., and Anandkumar, A.
\newblock Fourier neural operator for parametric partial differential
  equations.
\newblock \emph{arXiv preprint arXiv:2010.08895}, 2020{\natexlab{a}}.

\bibitem[Li et~al.(2020{\natexlab{b}})Li, Kovachki, Azizzadenesheli, Liu,
  Bhattacharya, Stuart, and Anandkumar]{li2020neural}
Li, Z., Kovachki, N., Azizzadenesheli, K., Liu, B., Bhattacharya, K., Stuart,
  A., and Anandkumar, A.
\newblock Neural operator: Graph kernel network for partial differential
  equations.
\newblock \emph{arXiv preprint arXiv:2003.03485}, 2020{\natexlab{b}}.

\bibitem[Liang et~al.(2022)Liang, Cao, Fan, Zhang, Ranjan, Li, Timofte, and
  Van~Gool]{liang2022vrt}
Liang, J., Cao, J., Fan, Y., Zhang, K., Ranjan, R., Li, Y., Timofte, R., and
  Van~Gool, L.
\newblock Vrt: A video restoration transformer.
\newblock \emph{arXiv preprint arXiv:2201.12288}, 2022.

\bibitem[Liu et~al.(2019)Liu, Ott, Goyal, Du, Joshi, Chen, Levy, Lewis,
  Zettlemoyer, and Stoyanov]{liu2019roberta}
Liu, Y., Ott, M., Goyal, N., Du, J., Joshi, M., Chen, D., Levy, O., Lewis, M.,
  Zettlemoyer, L., and Stoyanov, V.
\newblock Roberta: A robustly optimized bert pretraining approach.
\newblock \emph{arXiv preprint arXiv:1907.11692}, 2019.

\bibitem[Liu et~al.(2021)Liu, Lin, Cao, Hu, Wei, Zhang, Lin, and
  Guo]{liu2021swin}
Liu, Z., Lin, Y., Cao, Y., Hu, H., Wei, Y., Zhang, Z., Lin, S., and Guo, B.
\newblock Swin transformer: Hierarchical vision transformer using shifted
  windows.
\newblock In \emph{Proceedings of the IEEE/CVF International Conference on
  Computer Vision}, pp.\  10012--10022, 2021.

\bibitem[Mirocha et~al.(2014)Mirocha, Kosovi{\'c}, and
  Kirkil]{mirocha2014resolved}
Mirocha, J., Kosovi{\'c}, B., and Kirkil, G.
\newblock Resolved turbulence characteristics in large-eddy simulations nested
  within mesoscale simulations using the weather research and forecasting
  model.
\newblock \emph{Monthly Weather Review}, 142\penalty0 (2):\penalty0 806--831,
  2014.

\bibitem[Mishra et~al.(2019)Mishra, Mukhopadhaya, Iaccarino, and
  Alonso]{mishra2019uncertainty}
Mishra, A.~A., Mukhopadhaya, J., Iaccarino, G., and Alonso, J.
\newblock Uncertainty estimation module for turbulence model predictions in
  su2.
\newblock \emph{AIAA Journal}, 57\penalty0 (3):\penalty0 1066--1077, 2019.

\bibitem[Pathak et~al.(2020)Pathak, Mustafa, Kashinath, Motheau, Kurth, and
  Day]{pathak2020using}
Pathak, J., Mustafa, M., Kashinath, K., Motheau, E., Kurth, T., and Day, M.
\newblock Using machine learning to augment coarse-grid computational fluid
  dynamics simulations.
\newblock \emph{arXiv preprint arXiv:2010.00072}, 2020.

\bibitem[Pfaff et~al.(2020)Pfaff, Fortunato, Sanchez-Gonzalez, and
  Battaglia]{pfaff2020learning}
Pfaff, T., Fortunato, M., Sanchez-Gonzalez, A., and Battaglia, P.~W.
\newblock Learning mesh-based simulation with graph networks.
\newblock \emph{arXiv preprint arXiv:2010.03409}, 2020.

\bibitem[Pitsch(2006)]{pitsch2006large}
Pitsch, H.
\newblock Large-eddy simulation of turbulent combustion.
\newblock \emph{Annu. Rev. Fluid Mech.}, 38:\penalty0 453--482, 2006.

\bibitem[Ronneberger et~al.(2015)Ronneberger, Fischer, and
  Brox]{ronneberger2015u}
Ronneberger, O., Fischer, P., and Brox, T.
\newblock U-net: Convolutional networks for biomedical image segmentation.
\newblock In \emph{International Conference on Medical image computing and
  computer-assisted intervention}, pp.\  234--241. Springer, 2015.

\bibitem[Sanchez-Gonzalez et~al.(2020)Sanchez-Gonzalez, Godwin, Pfaff, Ying,
  Leskovec, and Battaglia]{sanchez2020learning}
Sanchez-Gonzalez, A., Godwin, J., Pfaff, T., Ying, R., Leskovec, J., and
  Battaglia, P.
\newblock Learning to simulate complex physics with graph networks.
\newblock In \emph{International Conference on Machine Learning}, pp.\
  8459--8468. PMLR, 2020.

\bibitem[Stachenfeld et~al.(2021)Stachenfeld, Fielding, Kochkov, Cranmer,
  Pfaff, Godwin, Cui, Ho, Battaglia, and
  Sanchez-Gonzalez]{stachenfeld2021learned}
Stachenfeld, K., Fielding, D.~B., Kochkov, D., Cranmer, M., Pfaff, T., Godwin,
  J., Cui, C., Ho, S., Battaglia, P., and Sanchez-Gonzalez, A.
\newblock Learned coarse models for efficient turbulence simulation.
\newblock \emph{arXiv preprint arXiv:2112.15275}, 2021.

\bibitem[Stone et~al.(2020)Stone, Tomida, White, and Felker]{Stone2020}
Stone, J.~M., Tomida, K., White, C.~J., and Felker, K.~G.
\newblock The athena++ adaptive mesh refinement framework: Design and
  magnetohydrodynamic solvers.
\newblock \emph{The Astrophysical Journal Supplement Series}, 249\penalty0
  (1):\penalty0 4, June 2020.
\newblock \doi{10.3847/1538-4365/ab929b}.
\newblock URL \url{https://doi.org/10.3847%2F1538-4365%2Fab929b}.

\bibitem[Vaswani et~al.(2017)Vaswani, Shazeer, Parmar, Uszkoreit, Jones, Gomez,
  Kaiser, and Polosukhin]{vaswani2017attention}
Vaswani, A., Shazeer, N., Parmar, N., Uszkoreit, J., Jones, L., Gomez, A.~N.,
  Kaiser, {\L}., and Polosukhin, I.
\newblock Attention is all you need.
\newblock In \emph{Advances in neural information processing systems}, pp.\
  5998--6008, 2017.

\bibitem[Wang et~al.(2020)Wang, Kashinath, Mustafa, Albert, and
  Yu]{wang2020towards}
Wang, R., Kashinath, K., Mustafa, M., Albert, A., and Yu, R.
\newblock Towards physics-informed deep learning for turbulent flow prediction.
\newblock In \emph{Proceedings of the 26th ACM SIGKDD International Conference
  on Knowledge Discovery \& Data Mining}, pp.\  1457--1466, 2020.

\end{thebibliography}
\bibliographystyle{icml2022}

\clearpage
\newpage
\appendix
\renewcommand\thefigure{\thesection.\arabic{figure}}  
\renewcommand{\thetable}{A\arabic{table}}

\section{Ablation Studies}
\setcounter{figure}{0} 
\setcounter{table}{0}
\label{sec:result-ablation}
We inspect the performances of the model when a certain component is removed. Specifically, one of Transformer's encoder, TMSA, and temporal embedding is deactivated while other parts of the architecture remain unchanged. We perform separate fine-tuning procedures, as described in Section \ref{sec:fine-tuning}, for each of the 'new' architecture. Table \ref{ablaiton-result} summarizes performances of this ablation study. We observe significant decrease in all four metrics when any of the three components is removed. Without temporal embedding, the model may lose important information on the temporal scale and unable to generate accurate predictions in longer rollout steps. This is further confirmed by the trend shown in Figure \ref{fig:ablation} with progressive increases in RMSE and decreaess in $\text{R}^2$ throughout the inference. In the absence of TMSA, dependencies between volumes at different temporal and spatial scales may be largely lost, resulting in a immediate performance reduction, as seen in Figure \ref{fig:ablation}. The removal of the Transformer encoder appears to fall between the prior two cases where portions of spatial and temporal information are lost. We conclude that every layer in our proposed architecture is necessary and functions as expected in capturing the dependencies in and between temporal and spatial scales.

\begin{table}[h]
\caption{Ablation studies on Transformer Encoder, TMSA, and temporal embedding}
\label{ablaiton-result}
\vskip 0.15in
\begin{center}
\begin{small}
\begin{sc}
\begin{tabular}{lcccr}
\toprule
Model & RMSE & MAE & $\text{R}^2$ & EV \\
\midrule
TNT & 0.0782  & 0.0561 & 0.8080 & 0.8239 \\
$\times$ Encoder & 0.1396 & 0.0998 & 0.3894 & 0.4066 \\
$\times$ Tmsa & 0.1371 & 0.1006 & 0.4073 & 0.4115 \\
\begin{tabular}{@{}c@{}}$\times$ Temporal \\ Embedding\end{tabular} & 0.1572 & 0.1195 & 0.2253 & 0.6056 \\
\bottomrule
\end{tabular}
\end{sc}
\end{small}
\end{center}
\vskip -0.1in
\end{table}

\begin{figure}[h]
\vskip 0.1in
\begin{center}
\centerline{\includegraphics[width=\columnwidth]{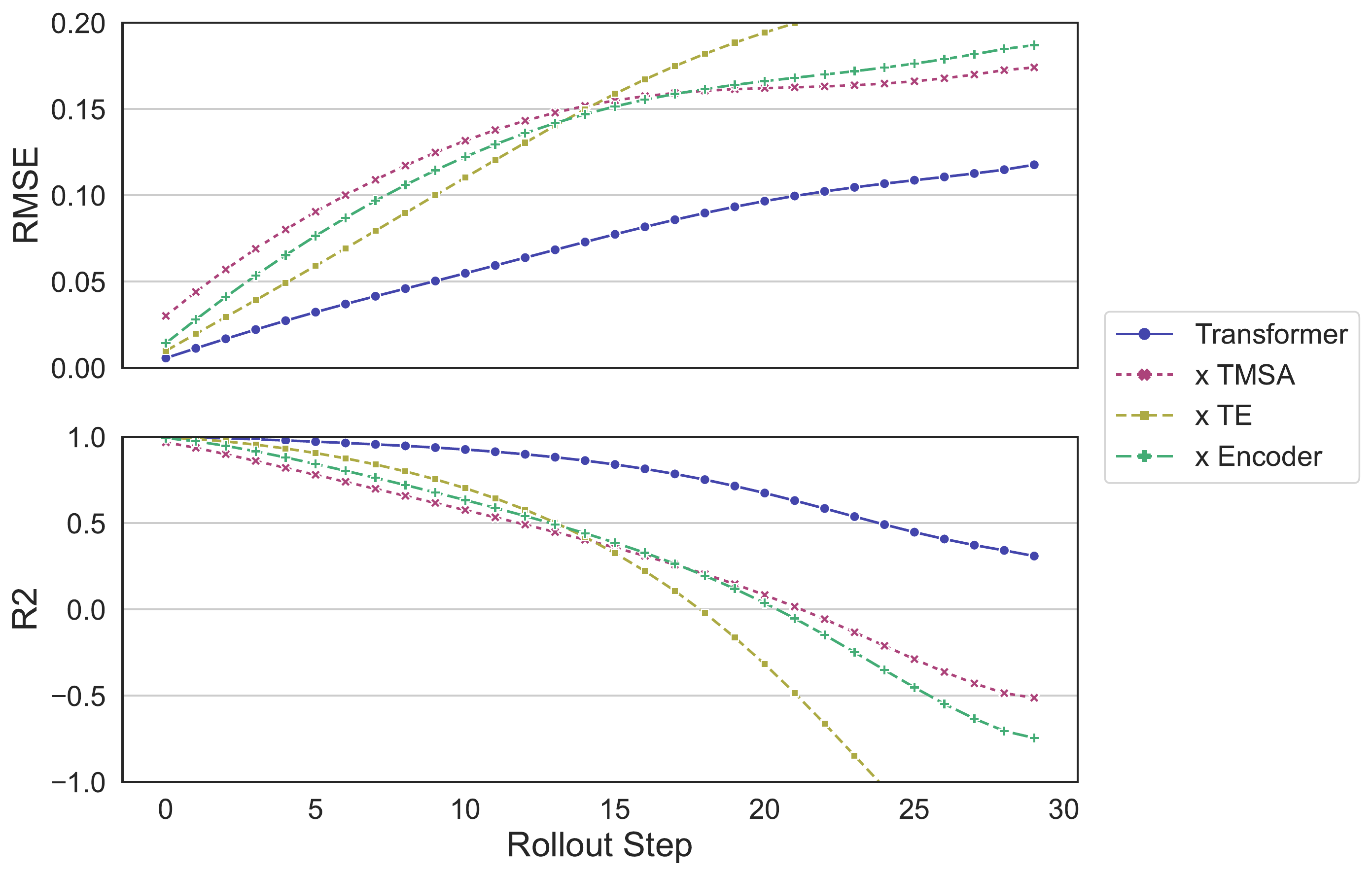}}
\caption{Ablation studies on Transformer Encoder, TMSA, and temporal embedding. RMSE and $\text{R}^2$ drop when any component is removed from the architecture.}
\label{fig:ablation}
\end{center}
\vskip -0.1in
\end{figure}

\section{Predictions on all features}
\setcounter{figure}{0} 
\setcounter{table}{0}
\label{sec:result-all-feature}
We further provide visual results in Figure \ref{fig:all_feature}. It shows slice plots on predictions of all five features at the end (step 30) of the rollout. It is clear that TNT stays robust even at long-range rolling prediction.
\begin{figure}[h]
\vskip 0.1in
\begin{center}
\centerline{\includegraphics[width=\columnwidth]{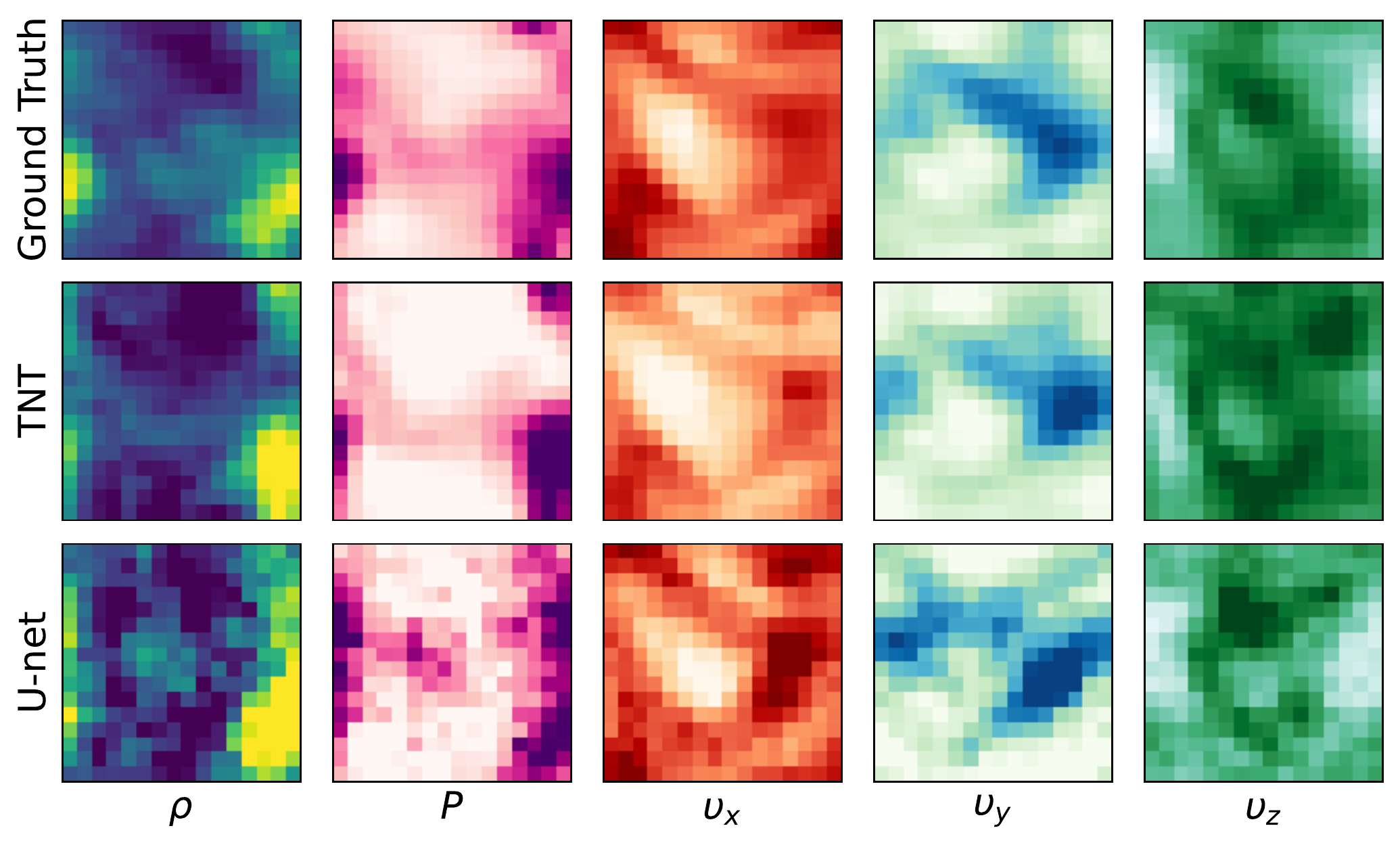}}
\caption{Visualization of predictions on all features at 30th rollout step of TNT and U-net vs ground truth. Each slice is taken at the middle of the volume across the third spatial axis $z$.}
\label{fig:all_feature}
\end{center}
\vskip -0.1in
\end{figure}

\section{Definition of Evaluation Metrics}
\setcounter{figure}{0}  
\setcounter{table}{0}

\label{sec:B}
Denote $y$ to be the target variable we want to predict with size $n$, and $\hat{y}$ to be the model predictions. Suppose $\hat{y_i}$ is the $i$-th prediction with $y_i$ the corresponding ground truth, then the evaluation metrics Root Mean Squared Error (RMSE), Mean Absolute Error (MAE), R-Squared ($\text{R}^2$), and Explained Variance (EV) are defined as follows:

\begin{itemize}
    \item $\text{RMSE}(y, \hat{y}) = \sqrt{\frac{1}{n} \sum_{i=0}^{n - 1} (y_i - \hat{y}_i)^2}$
    \item $\text{MAE}(y, \hat{y}) = \frac{1}{n} \sum_{i=0}^{n-1} \left| y_i - \hat{y}_i \right|$
    \item $\text{R}^2(y, \hat{y}) = 1 - \frac{\sum_{i=1}^{n} (y_i - \hat{y}_i)^2}{\sum_{i=1}^{n} (y_i - \bar{y})^2}$, where $\bar{y} = \frac{1}{n} \sum_{i=1}^{n} y_i$
    \item $\text{EV}(y, \hat{y}) = 1 - \frac{Var\{ y - \hat{y}\}}{Var\{y\}}$
\end{itemize}

RMSE calculates the expected value of the root of the squared loss or $l2$-norm loss, while MAE calculates the expected value of the absolute error loss or $l1$-norm loss. $\text{R}^2$ and EV represents the proportion of variance that has been explained by the model, and provides an indication of how well unseen samples are likely to be predicted by the model.

\end{document}